\documentclass[a4paper]{iopconfser}
\usepackage{graphicx}
\usepackage{amsmath,amssymb,float,subfigure}
\usepackage[document]{ragged2e}
\usepackage[makeroom]{cancel}
\usepackage{hyperref}
\usepackage[dvipsnames,svgnames,x11names,array,empheq,tikz,amsthm,multirow,amsfonts,ulem]{xcolor}
\usepackage[markup=underlined]{changes}
\begin{document}
\justifying
\title{Transverse envelope dynamics of beam slices in a uniform charged ellipsoidal model of the plasma bubble regime}

\author{Abdul Mannan$^{1,2}$, Alessio Del Dotto$^{1}$, and Massimo Ferrario$^{1}$}

\affil{$^1$Laboratori Nazionali di Frascati, INFN, Via Enrico Fermi 54, 00044, Frascati, RM, Italy}
\affil{$^2$Department of Physics, Jahangirnagar University, Savar, Dhaka-1342, Bangladesh}

\email{abdulmannan@juniv.edu}
\justifying
\begin{abstract}
We consider a pair of driver/witness electron bunches propagating in an ionized gas background
a configuration similar to the one produced in a capillary discharge where a plasma oscillation has
been excited by a driving pulse. We assume as in the plasma nonlinear regime that the plasma electrons behind the driver are
completely expelled and an ellipsoidal cavity filled with ions only is formed. The fields are linear
in both longitudinal and transverse directions, at least in the region of interest for particle acceleration, as
the one produced by a uniform ion distribution within a uniformly charged ellipsoidal volume. The fields
produced by the ions and experienced by a witness electron beam are purely electrostatic, being the ions at
rest in the laboratory frame on the time scale of interest and it can be represented with the field distribution
produced by a 3D charged ellipsoidal. The energy spread and emittance degradation has been studied by
slicing the bunch in an array of cylinders and solving envelope equations for each bunch slice. The properties
of transverse envelope and emittance oscillations and energy spread degradation have been analyzed together
with the related matching conditions for optimal transport and acceleration.
\end{abstract}

\section{Introduction}
It is well-known that a plasma wave having a phase velocity close to the speed of light $c$, as the one required in modern plasma-based accelerators, can be excited by a laser pulse \cite{Tajima1979} or by a particle beam \cite{Chen1985}. In the plasma wakefield excitation, the large amplitude accelerating fields are generated in the plasma when a relatively intense relativistic charged particle bunch \cite{Chen1985,Rosenzweig1988} (hereafter called the driver) is therein propagating. The main effect predicted by the theory, consistent with the experimental observations, is the excitation of a large amplitude plasma wave behind the driver. A witness bunch follows the driver at almost the same speed as the latter. The plasma wakefield excitation has been shown feasible in both overdense (the beam density is much smaller than the plasma density) and underdense (the beam density is nearly equal or a little above the plasma density) regimes. To describe the plasma wakefield excitation most of the theories are confined to linear fluid theory and one-dimensional (1D) nonlinear fluid theory. Recent plasma wakefield acceleration and laser wakefield acceleration experiments prove that the wakes produced in the bubble regime or so-called blowout regime cannot be described by 1D fluid theory. The plasma electrons behind the driver are completely expelled and a cavity filled with ions is formed. These expelled electrons make a narrow sheath outside the cavity filled with ions. The ions pull the expelled electrons back and overshoot. As a result, the plasma wave wake is produced. Nowadays, most of the laser-driven and beam-driven plasma wakefield accelerator experiments are performed in the bubble regime. Rosenweig \textit{et al.} \cite{Rosenweig1991} first investigated the blowout regime of the plasma wakefield accelerator for high-intensity electron beam. Pukhov and Meyer-Ter-Vehn \cite {Pukhov2002} observed that the ion channel has a solitary plasma cavity shape for a highly intense laser driver. Lu \textit{et al.} \cite{Lu2006} presented a nonlinear theory for wake excitation by laser driver or particle beam. They showed that the ion channel is spherical shaped for large blowout radius. It is also reported that ion column has an ellipse shape for the blowout radius ($1<r_m<4$). The bubble shape and electromagnetic fields for ultra-intense laser wakefield acceleration have been studied by Li \textit{et al.} \cite{Li2015}. They assumed the bubble shape to be ellipsoidal where the bubble moves with the speed of light and the residual electrons inside the bubble were ignored.

When longitudinal correlations within the bunch are important like the one induced by space charge, longitudinal energy spread or any position dependent external fields, beam envelope evolution is generally dependent also on the coordinate along the bunch. In this case, the bunch should be considered as an ensemble of $n$ longitudinal slices, whose evolution can be modelled by $n$-slice envelope equations, provided that the bunch parameters refer to each single slice. Correlations within the bunch may cause emittance degradation that can be evaluated, once an analytical or numerical solution of the slice envelope equation is known. The basic approximation in the description of beam dynamics \cite{Ferrario2020} lies in the assumption that each bunch is described by a uniformly charged cylinder (with circular or elliptical cross-section) whose time evolution is laminar both in transverse and longitudinal directions. Under these circumstances, envelope equations can be conveniently adopted to describe the bunch dynamics. By slicing the bunch in an array of cylinders so-called Multi-slices approximation \cite{Ferrario2007}, each one subject to the local field, one obtains the energy spread and the emittance degradation.

In this paper, we study the dynamics of beam envelopes in the elliptical blowout regime. We assume that an ellipsoid blowout regime or bubble filled by ions is formed when the driver passes through the plasma. To describe the rms driver/witness beam envelope dynamics we derive coupled envelope equations. The equilibrium solution (i.e. the condition for matching beams in a plasma accelerator) of coupled envelope equations under some assumptions is found. Finally, the numerical and simulation results are compared and discussed.

\section{Ellipsoidal Bubble Model}
\begin{figure}[H]%
\centering
\includegraphics[width=100mm]{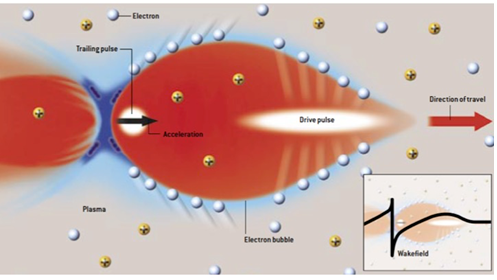}
\caption{Schematic representation of the longitudinal wake field (black line) and ion distribution (red area) behind a driving laser or particle beam \cite{Joshi2006}.}\label{Fig1}
\end{figure}

In the non-linear regime the plasma electrons behind the driver are almost completely expelled and an ellipsoidal cavity filled by ions only is formed. A very simplified model for the plasma behind the driving pulse is illustrated in Figure \ref{Fig1}. We have considered an ellipsoidal uniform ion distribution, indicated by the dashed ellipsoid, with particle density $n_p$. We have focused in this section on the beam driven case, i.e. a plasma wave excited by a driving electron bunch with particle density $n_b$, but the model can be easily extended to describe also the laser driven case. This model is justified by the fact that in this regime the fields are linear in both longitudinal and transverse directions, at least in the region of interest for particle acceleration, as the one produced by a uniform ion distribution within an ellipsoid of semi-axes:
\begin{equation}\label{axes}
\left.
\begin{aligned}
X&=6.32\sqrt{\alpha}\,\sigma_{xd}\\
Y&=6.32\sqrt{\alpha}\,\sigma_{yd}\\
Z&=\frac{\lambda_p}{2}
\end{aligned}
\right\}
\end{equation}
where $\alpha = n_b/n_p\geq 1$ in the nonlinear regime, $\sigma_{xd}$ and $\sigma_{yd}$ represent the driving beam spot size and $\lambda_p = 2\pi/k_p$ is the plasma wavelength. The computation of the $X$ and $Y$ semi-axes has been derived in Ref. \cite{Lu2006} while the assumption on $Z$ is a reasonable approximation.
\par
The fields produced by the ions and experienced by a witness electron beam is purely electrostatic, being the ions at rest in the laboratory frame on the time scale of interest and it can be represented with the field distribution produced by a 3D ellipsoidal filled by a charge of $Q_B = 4\pi XYZen_p/3$ according to the expressions derived by Lapostolle \cite{Lapostolle1965} and reported also in \cite{Wangler1998}:
\begin{equation}\label{be-field}
\left.
\begin{aligned}
E_x&=\frac{3Q_B(1-f_d)}{4\pi\varepsilon_0(X+Y)Z}\frac{x}{X} = \frac{en_p}{\varepsilon_0}\frac{Y}{X+Y}(1-f_d)x\\
E_y&=\frac{3Q_B(1-f_d)}{4\pi\varepsilon_0(X+Y)Z}\frac{y}{Y} = \frac{en_p}{\varepsilon_0}\frac{X}{X+Y}(1-f_d)y\\
E_z&=\frac{3Q_Bf_d}{4\pi\varepsilon_0 XY}\frac{z}{Z} = \frac{en_p}{\varepsilon_0}f_dz
\end{aligned}
\right\}
\end{equation}
The quantity $f_d$ is the ellipsoid form factor that can be defined as:
\begin{equation}\label{formfactor}
 f_d=\frac{1}{3}\frac{\sqrt{XY}}{Z} = 0.21\sqrt{\frac{Q_dr_e}{e\sigma_{zd}}}\,.
\end{equation}

The electric field $E_z$ is moving along the longitudinal axis, due to the plasma electron collective oscillation, with the speed of the driving pulse, even if the source of the field, i.e. the ions remains at rest in the laboratory frame. The propagating electric field through the plasma indeed produces a time-varying azimuthal magnetic field $B_\vartheta$. From the Amp\'ere's law, one can compute the expression of the induced magnetic field:
\begin{equation}\label{B}
  B_x=-\frac{en_p}{2\varepsilon_0c}f_d y ~~~~\text{and}~~~~ B_y=\frac{en_p}{2\varepsilon_0c}f_d x\,.
\end{equation}
The accelerating component of the field is linearly increasing from the moving ellipsoidal bubble center and depends on both the plasma density and the driver parameters. By changing the longitudinal coordinate $\zeta=z-z_c$ and using the definition of $X$, $Y$ and $f_d$ we can write for the accelerating field:
\begin{equation}\label{Ezeta}
  E_z(\zeta)= \frac{en_p}{\varepsilon_0}f_d\zeta = \frac{5.95e}{\varepsilon_0}\frac{\sqrt{n_pn_d\sigma_{xd}\sigma_{yd}}}{\lambda_p}\zeta\,.
\end{equation}
If the driving pulse has a uniform cylindrical charge distribution the particle density can be written as:
\begin{equation}\label{nd}
  n_d = \frac{Q_d}{\pi e(2\sqrt{3})^{3}\sigma_{xd}\sigma_{yd}\sigma_{zd}} = \frac{I_d}{12\pi e c\sigma_{xd}\sigma_{yd}}
\end{equation}
with peak current $I_d = Q_d/(2\sqrt{3}\sigma_t)$. The resulting accelerating field along the bubble axis reduces to:
\begin{equation}\label{Ezeta1}
  E_z(\zeta)=An_p\sqrt{I_d}\zeta\,,
\end{equation}
where the constant $A= \frac{1.48}{\sqrt{3m_0}}(\frac{e}{\pi\varepsilon_0c})^{3/2}$.

The maximum gradient occurs on the bubble edge at $\zeta = \lambda_p/2$ and scales like $E_z(\lambda_p/2) = \tilde{A} \sqrt{n_pI_d}$ with the constant $\tilde{A} = \frac{1.48}{\varepsilon_0}\sqrt{\frac{e}{3\pi c}}$.

Using the fields [as expressed in \eqref{be-field} and \eqref{B}], the Lorentz force acting on the electrons can be written as:
\begin{equation}\label{Fx}
  F_x = \frac{k_p^2m_0c^2}{2}\frac{\left(2\sigma_{yd}-
  (\sigma_{xd}+3\sigma_{yd})f_d\right)}{\sigma_{xd}+\sigma_{yd}}x
\end{equation}
\begin{equation}\label{Fy}
  F_y = \frac{k_p^2m_0c^2}{2}\frac{\left(2\sigma_{xd}-
  (\sigma_{yd}+3\sigma_{xd})f_d\right)}{\sigma_{xd}+\sigma_{yd}}y\,.
\end{equation}
Here, $k_p^2 = e^2n_p/(\varepsilon_0m_0c^2)$ is the plasma wave number.
The equations for the longitudinal centroid motion for each slice of the bunch traveling on the axis are represented by the following differential
equation:
\begin{equation}\label{zc}
  \frac{dz_s}{dt}=\beta_s c~~~~~\text{and}~~~~~ \frac{d\beta_s}{dt} = \frac{e}{m_0c\gamma_s^3}E_z(\zeta,t)
\end{equation}
where the expression of the lorentz factor for each slice $s$ is given by $\gamma_s=1/(1-\beta_s^2)^{1/2}$.

It is worth mentioning that the total electric field acting on the trailing or witness beam is given by
\begin{equation}\label{total-electric}
  E_{z,t} = E_d(z) - \mathcal{T} E_w(z) = \frac{en_p}{\varepsilon_0}\left(f_d - \mathcal{T}f_w\right)z\,,
\end{equation}
where $\mathcal{T}$, which is equal to $1/2$, accounts for the fundamental theorem of beam loading \cite{Chao2013,Pompili2022} and the $f_w$ is given by
\begin{equation}\label{formfactor-w}
 f_w=0.21\sqrt{\frac{Q_wr_e}{e\sigma_{zw}}}\,.
\end{equation}
\par
Note that we assume in our present investigation the driving/witness beam is a uniformly charged cylinder. The uniformly charged cylinder is divided into slices whose evolution is described by the envelope equations, like the beam. Therefore, one has to consider the average values of all slices' spot sizes to calculate the semi-axes (i.e., $X$ and $Y$) of the ellipsoid bubble. The average $\langle ~ \rangle = \frac{1}{N}\sum_{s=1}^N$ is performed over the $N$ slices. We also consider a uniform charge distribution within each slice. We finally obtain the complete coupled rms envelope equations for the slice of driver and witness as \cite{Ferrario2020,Sacherer1971,Rosenzweig2003,Reiser1994}:
\begin{equation}\label{envelope-x2}
  \sigma_{xd}'' + \frac{\gamma_d'}{\gamma_d}\sigma_{xd}' + \mathcal{T}\frac{k_p^2}{2\gamma_d}\frac{2\langle\sigma_{yd}\rangle-
  (\langle\sigma_{xd}\rangle+3\langle\sigma_{yd}\rangle)f_d}{\langle\sigma_{xd}\rangle+\langle\sigma_{yd}\rangle}\sigma_{xd} = \frac{\varepsilon_{nd}^2}{\gamma_d^2\sigma_{xd}^3}+\frac{2I_d}{I_A\gamma_d^3\sigma_{xd}}
\end{equation}
\begin{equation}\label{envelope-y2}
  \sigma_{yd}'' + \frac{\gamma_d'}{\gamma_d}\sigma_{yd}' +\mathcal{T}\frac{k_p^2}{2\gamma_d}\frac{2\langle\sigma_{xd}\rangle-
  (\langle\sigma_{yd}\rangle+3\langle\sigma_{xd}\rangle)f_d}{\langle\sigma_{xd}\rangle+\langle\sigma_{yd}\rangle}\sigma_{yd} = \frac{\varepsilon_{nd}^2}{\gamma_d^2\sigma_{yd}^3}+\frac{2I_d}{I_A\gamma_d^3\sigma_{yd}}
\end{equation}
\begin{multline}\label{envelope-wx2}
  \sigma_{xw}'' + \frac{\gamma_w'}{\gamma_w}\sigma_{xw}' + \frac{k_p^2}{2\gamma_w}\left(\frac{2\langle\sigma_{yd}\rangle-
  (\langle\sigma_{xd}\rangle+3\langle\sigma_{yd}\rangle)f_d}{\langle\sigma_{xd}\rangle+\langle\sigma_{yd}\rangle}+\mathcal{T}\frac{2\langle\sigma_{yw}\rangle-
  (\langle\sigma_{xw}\rangle+3\langle\sigma_{yw}\rangle)f_w}{\langle\sigma_{xw}\rangle+\langle\sigma_{yw}\rangle}\right)\sigma_{xw}\\
 = \frac{\varepsilon_{nw}^2}{\gamma_w^2\sigma_{xw}^3}+\frac{2I_w}{I_A\gamma_w^3\sigma_{xw}}
\end{multline}
\begin{multline}\label{envelope-wy2}
  \sigma_{yw}'' + \frac{\gamma_w'}{\gamma_w}\sigma_{yw}' + \frac{k_p^2}{2\gamma_w}\left(\frac{2\langle\sigma_{xd}\rangle-
  (\langle\sigma_{yd}\rangle+3\langle\sigma_{xd}\rangle)f_d}{\langle\sigma_{xd}\rangle+\langle\sigma_{yd}\rangle}+\mathcal{T}\frac{2\langle\sigma_{xw}\rangle-
  (\langle\sigma_{yw}\rangle+3\langle\sigma_{xw}\rangle)f_w}{\langle\sigma_{xw}\rangle+\langle\sigma_{yw}\rangle}\right)\sigma_{yw} \\
  = \frac{\varepsilon_{nw}^2}{\gamma_w^2\sigma_{yw}^3}+\frac{2I_w}{I_A\gamma_w^3\sigma_{yw}}
\end{multline}
where the alfv\'en current $I_A = 4\pi\varepsilon_0 m_0 c^3/e$ and the emittance $\varepsilon_{n}$ is normalized by $\gamma$.
\par
We now look for an equilibrium solution to the above-coupled equations under some assumptions. We neglect the adiabatic term ($\gamma'_{d,w}$ = 0), i.e. no acceleration.
Under this assumption, one can find a proper matching condition which holds at the entrance and exit of the plasma column. We finally obtain the equilibrium solution of \eqref{envelope-x2} - \eqref{envelope-wy2} with the above-mentioned condition as
\begin{equation}\label{solution-d}
  \sigma_{eq,xd,yd} = \left(\frac{2I_{d}+\sqrt{4I_{d}^2+2I_A^2\gamma_d^3k_p^2\varepsilon_{nd}^2(1-2f_d)\mathcal{T}}}{I_A\gamma_d^2k_p^2(1-2f_d)\mathcal{T}}\right)^{1/2}\,,
\end{equation}
\begin{equation}\label{solution-w}
  \sigma_{eq,xw,yw} =  \left(\frac{2I_{w}+\sqrt{4I_{w}^2+2I_A^2\gamma_w^3k_p^2\varepsilon_{nw}^2(1-2f_d+(1-2f_w)\mathcal{T})}}
  {I_A\gamma_w^2k_p^2(1-2f_d+(1-2f_w)\mathcal{T})}\right)^{1/2}\,.
\end{equation}
\begin{figure}[H]
\includegraphics[width=0.95\textwidth]{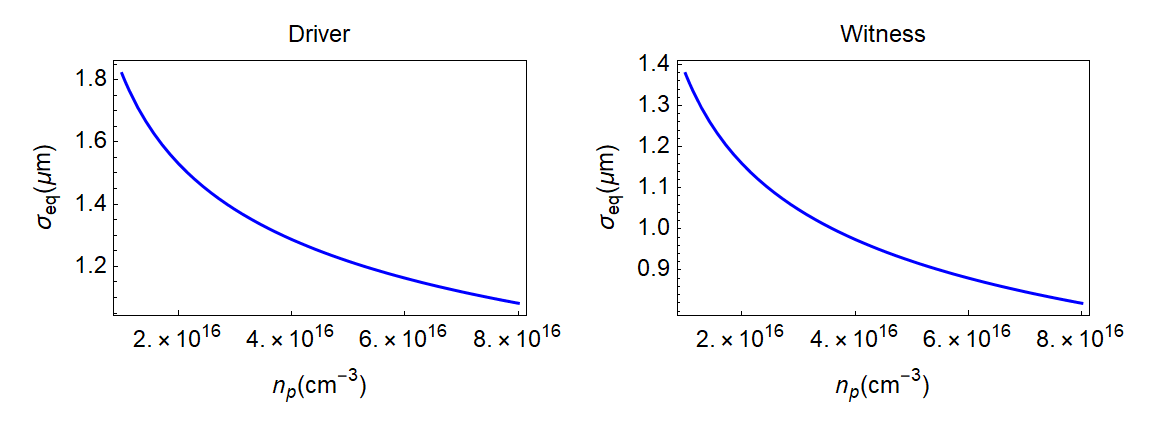}
\caption{The matched beam envelope [represented by the equilibrium solution of \eqref{solution-d} and \eqref{solution-w}]. The physical parameters of the driving beam (Left) and witness beam (Right) are displayed in Table \ref{tab1}.}
\label{equilibrium-sol}
\end{figure}
 The variation of the matched beam spot size [given by \eqref{solution-d} and \eqref{solution-w}] against plasma density, $n_p$ is displayed in figure \ref{equilibrium-sol}. The spot size of both driving and witness beams decreases with the increasing value of plasma density.
\section{Results and Discussion}
 Since an exact analytical solution of the coupled rms envelope equations [given by \eqref{envelope-x2} - \eqref{envelope-wy2}] is not possible we want to solve them numerically. For the numerical solution, we use the multi-slice approximation while the beam is traveling on the axis \cite{Ferrario2007}. We have used the following parameters: the Alfv\'en current 17 kA, the plasma density $10^{16}$ cm$^{-3}$, the capillary length $60$ cm, the ramp size $0.1$ cm, the number of slices of beam $N=150$, and the other parameters as shown in Table \ref{tab1}.

 We present here two examples for the (i) symmetric and (ii) asymmetric transverse beam spot size. We have numerically integrated equations \eqref{envelope-x2} - \eqref{envelope-wy2} from $z = 0$ to $z= \infty$ with the initial conditions (i) $\sigma_{xd}=\sigma_{yd} = 1.81 ~\mu$m (symmetric case), (ii) $\sigma_{xd}= 1.81~\mu$m; $\sigma_{yd} = 2.00 ~\mu$m (asymmetric case), $\sigma_{xd}'=\sigma_{yd}' =0$ at $z=0$; (i) $\sigma_{xw}=\sigma_{yw} = 1.37 ~\mu$m (symmetric case); (ii) $\sigma_{xw}= 1.37~\mu$m; $\sigma_{yw} = 1.50 ~\mu$m (asymmetric case), $\sigma_{xw}'=\sigma_{yw}' =0$ at $z=0$. For the mismatched asymmetric beam spot size, we choose the initial transverse spot size of both driver and witness as $3~\mu$m along the $x$-axis and the spot size in $y$-axis is greater than $10$\% of the spot size in $x$-axis. Note that the energy associated with both driving and trailing beams $\gamma_s$, which is a function of $z$, has been calculated for each slice with the numerical integration of \eqref{zc} with the initial condition.
 \par
 In order to evaluate the degradation of the rms emittance produced by longitudinal correlation in space charge effect and transverse external forces, we use the following expression for the correlated emittance \cite{Ferrario2007}
\begin{equation}\label{emittance}
  \varepsilon_{n,x}^\text{cor} = \sqrt{\langle \sigma_x^2 \rangle\langle (\gamma_s\sigma_x')^2\rangle - \langle\sigma_x\gamma_s \sigma_x'\rangle^2}\,,
\end{equation}
where the average $\langle ~ \rangle = \frac{1}{N}\sum_{s=1}^N$ is performed over the $N$ slices and an analogous equation holds for the $\sigma_y$ envelope.
The total rms emittance will be given by
\begin{equation}\label{emittance_t}
  \varepsilon_n = \sqrt{(\varepsilon_n^\text{th})^2+(\varepsilon_n^\text{cor})^2}\,.
\end{equation}
The energy spread is defined as \cite{Ferrario2007}:
\begin{equation}\label{energy-spread}
  \frac{\Delta \gamma}{\gamma} = \frac{\sqrt{\langle(\gamma_s - \langle\gamma_s\rangle)^2\rangle}}{\langle \gamma_s \rangle}\,.
\end{equation}
\begin{table}[h]
\caption{Physical parameters for driving and witness beams.}
\centering
\begin{tabular}{|l|p{3cm} |p{3cm}|}
\hline
&\textbf{Driver} & \textbf{Witness}\\
\hline
charge, $Q$ & 150 pC & 50 pC \\
\hline
length, $\sigma_z$& 30 $\mu$m &  20 $\mu$m\\
\hline
matched symmetric  & $\sigma_x = 1.81 ~ \mu$m & $\sigma_x =  1.37~ \mu$m\\

 spot size          & $\sigma_y = 1.81 ~ \mu$m & $\sigma_y = 1.37~ \mu$m\\
                  \hline
matched asymmetric   & $\sigma_x =1.81 ~ \mu$m & $\sigma_x = 1.37~ \mu$m\\

spot size             & $\sigma_y =2.0 ~ \mu$m & $\sigma_y = 1.50~ \mu$m\\
\hline
mismatched symmetric  & $\sigma_x = 3.0 ~ \mu$m & $\sigma_x =  3.0~ \mu$m\\

 spot size          & $\sigma_y = 3.0 ~ \mu$m & $\sigma_y = 3.0~ \mu$m\\
                  \hline
mismatched asymmetric   & $\sigma_x =3.0 ~ \mu$m & $\sigma_x = 3.0~ \mu$m\\

spot size             & $\sigma_y =3.3 ~ \mu$m & $\sigma_y = 3.3~ \mu$m\\
\hline
rms norm. emittance&  1.0 $\mu$m rad & 1.0 $\mu$m rad \\
\hline
energy  & 600 MeV & 600 MeV\\
\hline
\end{tabular}
\label{tab1}
\end{table}
\begin{figure}[H]
  \centering
   \begin{tabular}{@{}cc@{}}
   \includegraphics[width=0.48\textwidth]{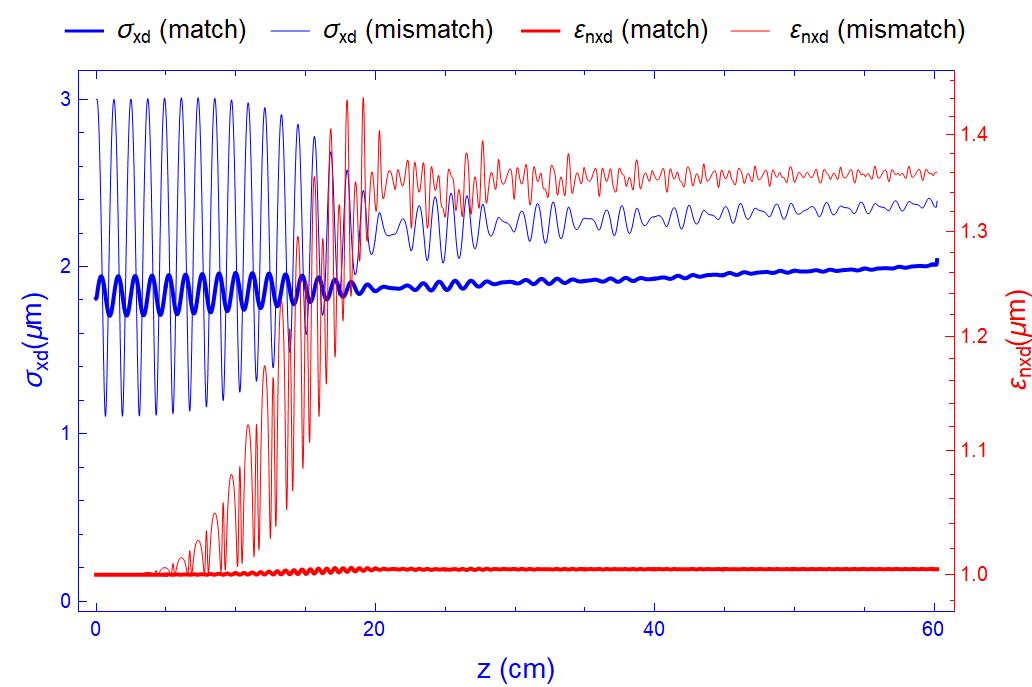} &
   \includegraphics[width=0.48\textwidth]{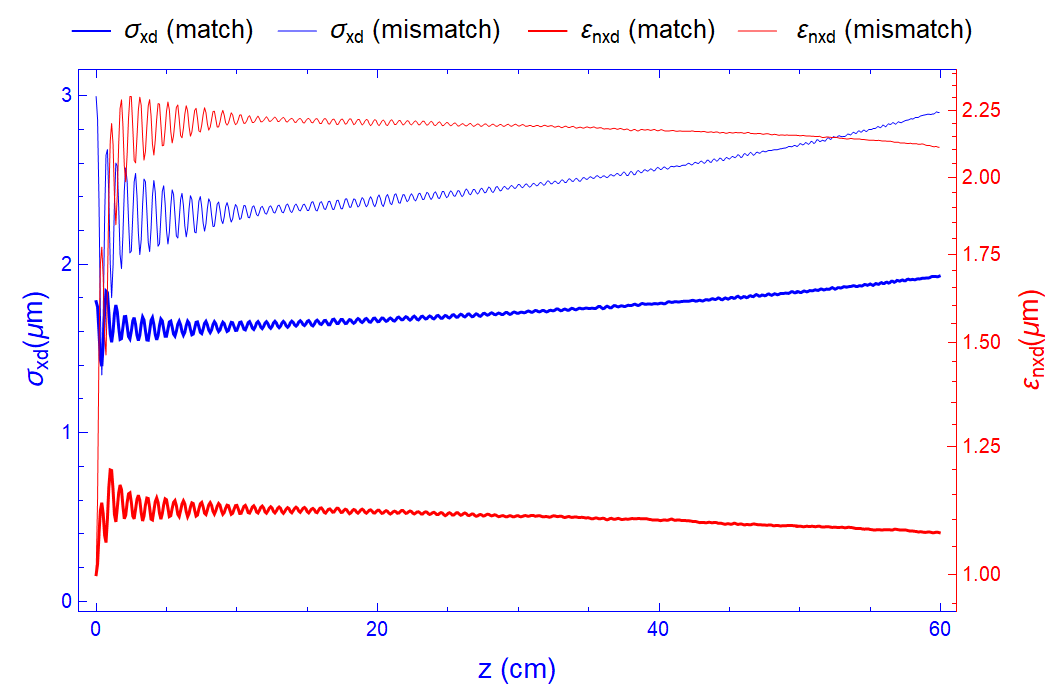}
\end{tabular}
  \caption{The evolution of transverse spot size and emittance in the case of driving beam: (i) numerical results (left side) and (ii) PIC simulation results (right side).}\label{spot-d}
\end{figure}
\begin{figure}[H]
  \centering
   \begin{tabular}{@{}cc@{}}
   \includegraphics[width=0.48\textwidth]{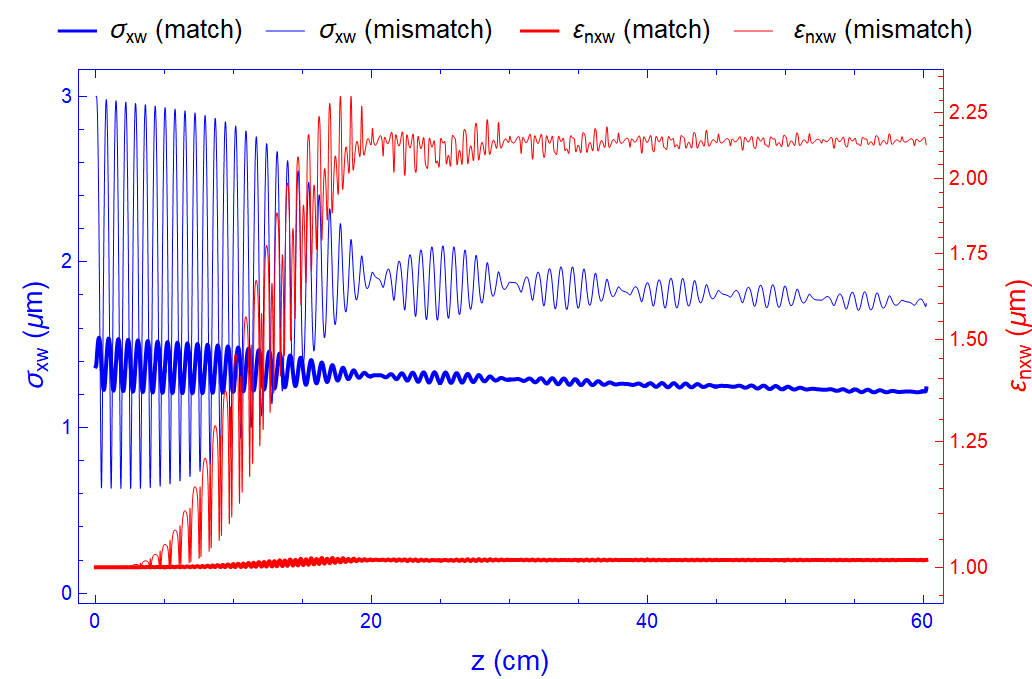} &
   \includegraphics[width=0.48\textwidth]{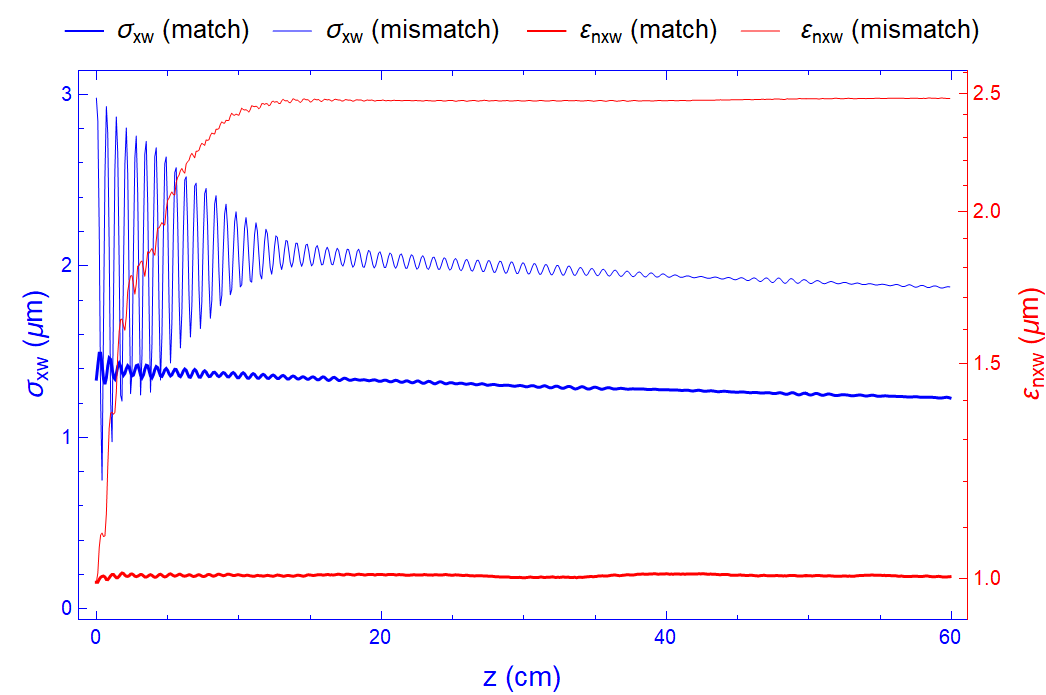}
\end{tabular}
  \caption{The evolution of transverse spot size and emittance in the case of trailing beam: (i) numerical results (left side) and (ii) PIC simulation results (right side).}\label{spot-w}
\end{figure}
\begin{figure}[H]
  \centering
  \begin{tabular}{@{}cc@{}}
   \includegraphics[width=0.48\textwidth]{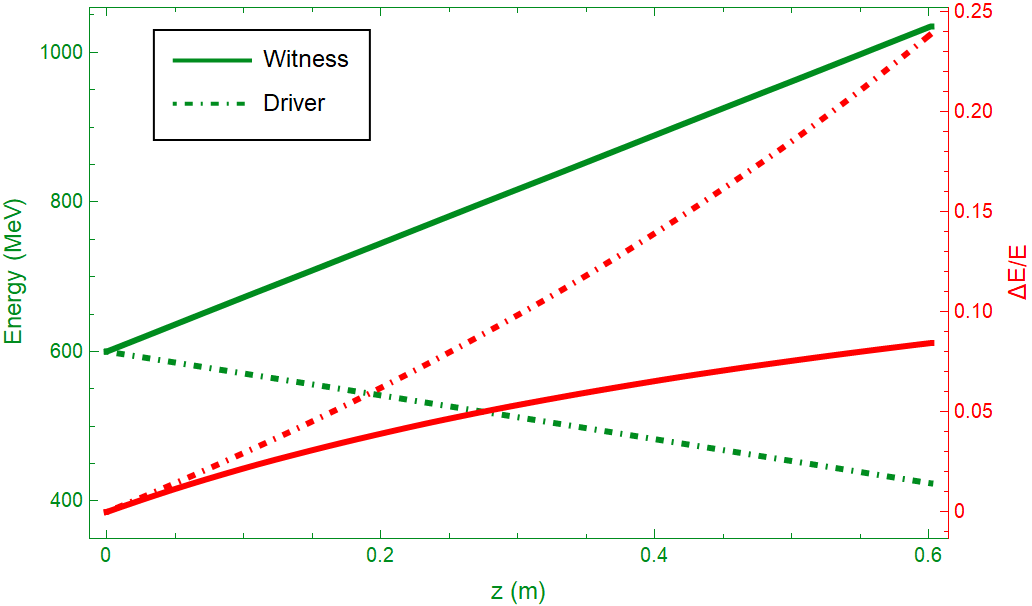} &
   \includegraphics[width=0.48\textwidth]{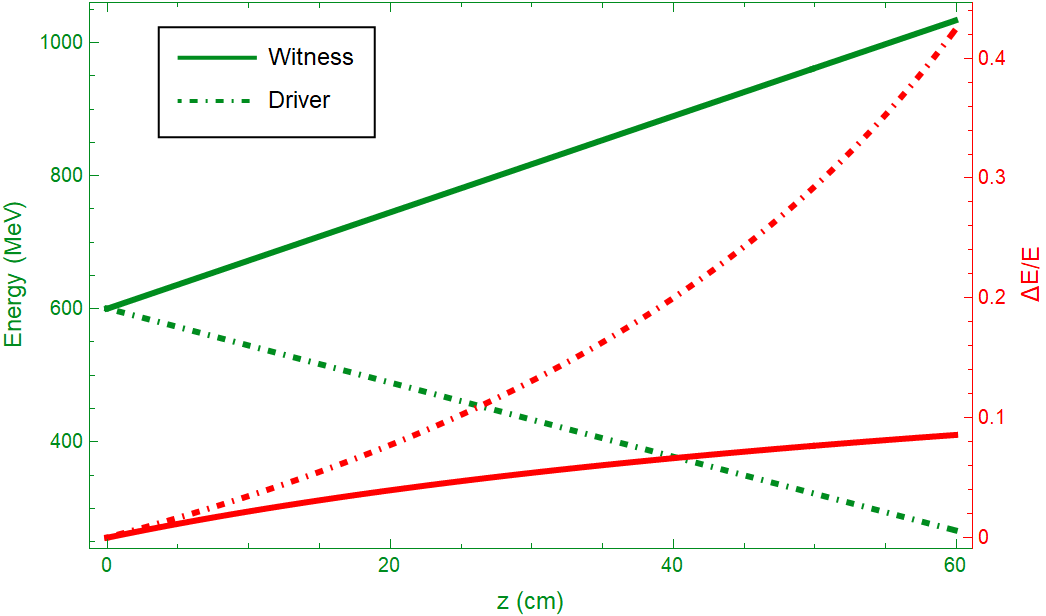}
\end{tabular}
  \caption{The energy and energy spread in the case of both driver and trailing beams: (i) numerical results (left side) and (ii) PIC simulation results (right side).}\label{fig-energy-spread}
\end{figure}
\begin{figure}[H]
\begin{tabular}{@{}cc@{}}
   \includegraphics[width=0.48\textwidth]{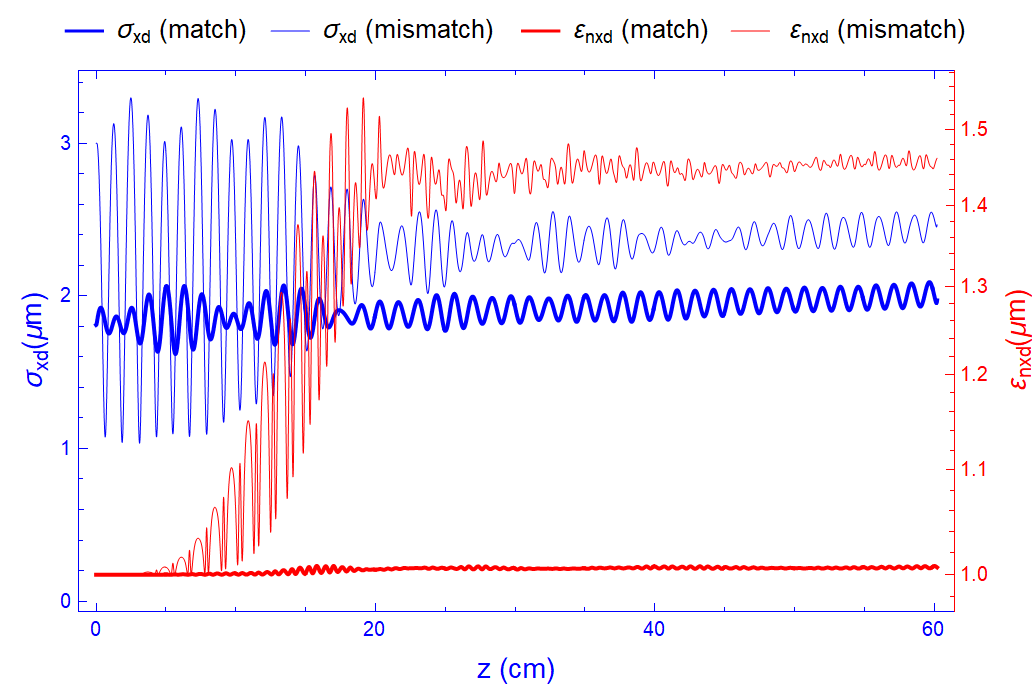} &
   \includegraphics[width=0.48\textwidth]{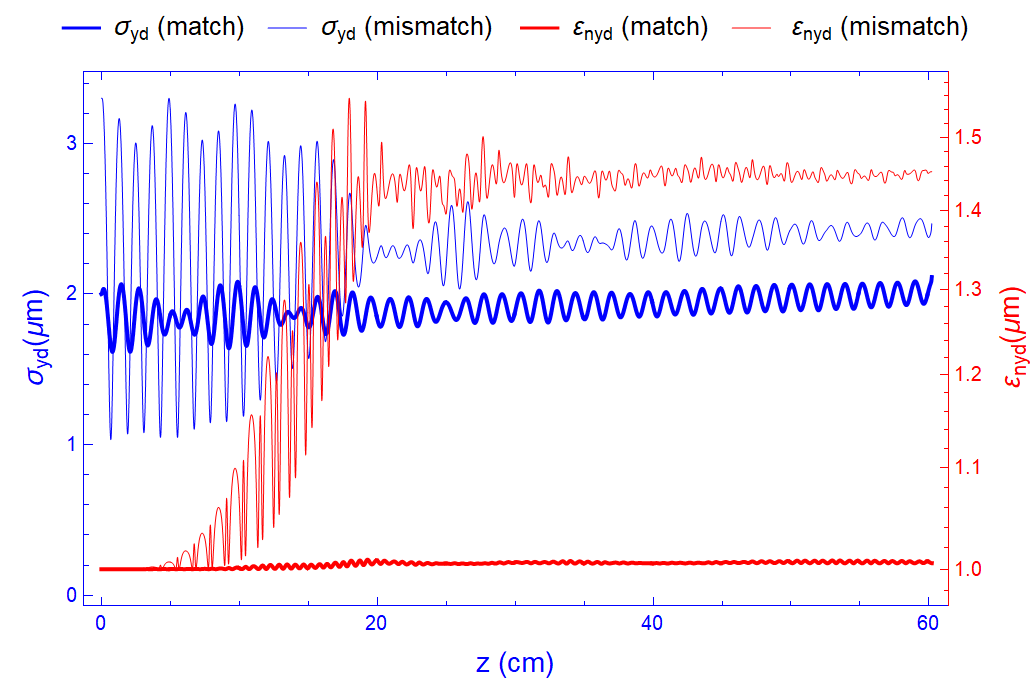}
\end{tabular}
\caption{The evolution of asymmetric transverse spot size and normalized emittance of driving beam.}
\label{ex-dx-dy}
\end{figure}
\begin{figure}[H]
\begin{tabular}{@{}cc@{}}
   \includegraphics[width=0.48\textwidth]{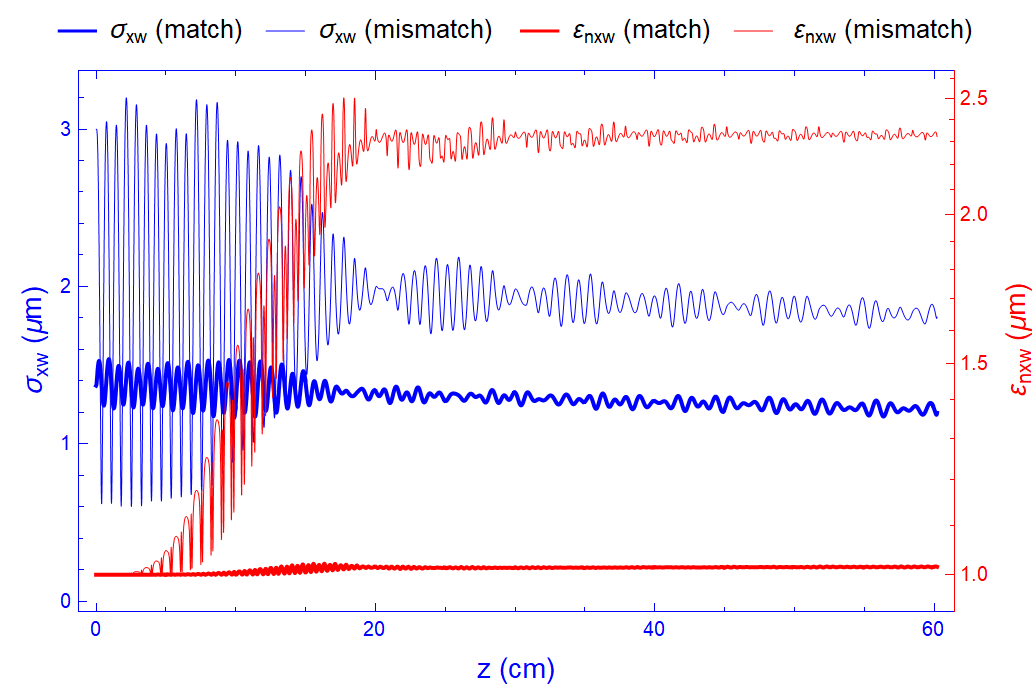} &
   \includegraphics[width=0.48\textwidth]{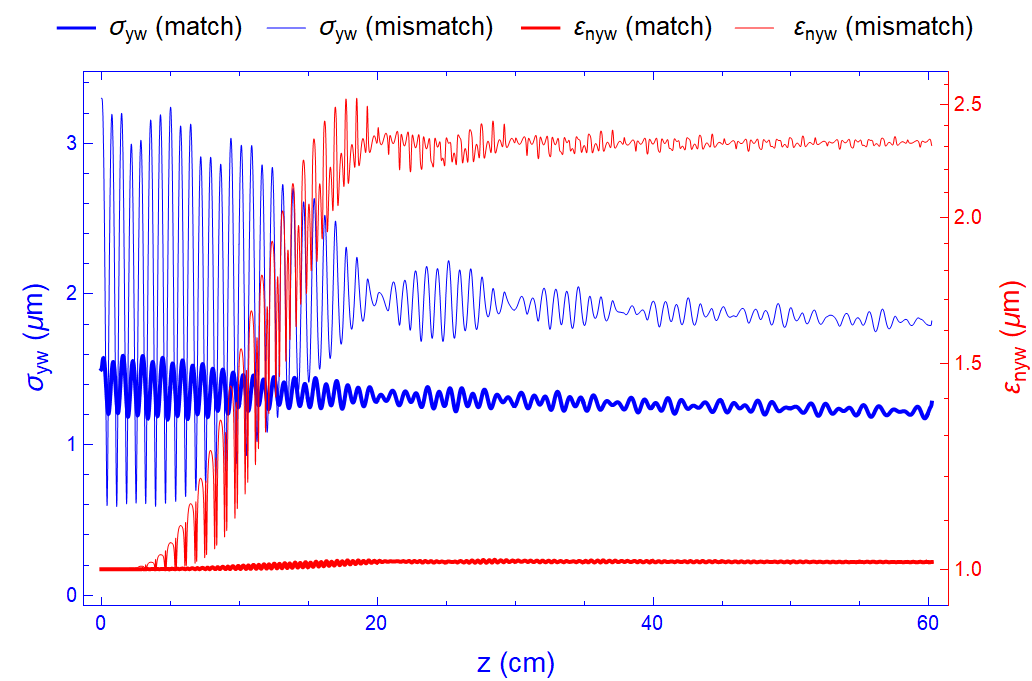}
\end{tabular}
\caption{The evolution of asymmetric transverse spot size and normalized emittance of witness beam.}
\label{ex-wx-wy}
\end{figure}
The numerical and particle-in-cell (PIC) simulation results are displayed in Figs. \ref{spot-d}-\ref{ex-wx-wy}. Figures \ref{spot-d} - \ref{ex-wx-wy} display the evolution of both driver and witness beams with symmetric and asymmetric transverse spot sizes. It is seen that both the transverse spot size and emittance saturate very quickly if we match the beam with the spot size which is given by \eqref{solution-d} and \eqref{solution-w}. However, the oscillation of transverse spot size and growth of emittance have been observed at the beginning in the case of mismatch beam and then reach saturation. The energy and energy spread of both driver and witness beams are shown in Fig. \ref{fig-energy-spread}. We have compared our numerical results with the PIC simulation results. The simulation results have been performed by using the hybrid code Architect \cite{Marocchino2016,Massimo2016} where we have used the same set of parameters (as shown in Table \ref{tab1}). It is found that our result has a good agreement with the PIC simulation results.

\section{Conclusions}
We have presented an elliptical model for the blowout regime in plasma wakefield acceleration. The model provides the longitudinal and transverse field produced by a driving bunch interacting with the plasma: the ion-filled region is modeled to have an elliptical shape and beam density greater than plasma density has been considered. Within that framework, we studied the dynamics of the driving bunch and a following witness trailing bunch. The envelope equations of the coupled system have been provided and solved, analytically for finding the matching conditions in simplified conditions, namely disregarding the adiabatic term and numerically in the general case. The length of the driver slice has been elongated since the driver decelerates itself. On the other hand, the slice length of the witness beam has been contracted since the witness beam feels the wakefield and is accelerated. As a result, the peak current of each slice of the driver (witness) decreases (increases). If a beam is matched to an adiabatic plasma profile, the transverse spot size and emittance oscillate around its initial value with a small amplitude. For the mismatch beam, the emittance growth and transverse spot size oscillation have been observed at the beginning and then they reach a stable condition. The properties of the transverse envelope and emittance oscillation for both symmetric and asymmetric driver/witness beams have been studied.
\section*{Acknowledgments}
Abdul Mannan gratefully acknowledges the financial support of the Istituto Nazionale di Fisica Nucleare - INFN (Italy) through its
post-doctoral research fellowship.

\end{document}